\documentclass[12pt]{article}

\usepackage{amssymb,amsmath,amsfonts,eurosym,geometry,ulem,graphicx,caption,color,setspace,sectsty,comment,footmisc,caption,natbib,pdflscape,subfigure,array,hyperref,booktabs,tabularx,float,tikz,authblk}
\usetikzlibrary{positioning, arrows.meta, calc}
\usepackage{xurl}
\newcolumntype{L}[1]{>{\raggedright\let\newline\\arraybackslash\hspace{0pt}}m{#1}}
\newcolumntype{C}[1]{>{\centering\let\newline\\arraybackslash\hspace{0pt}}m{#1}}
\newcolumntype{R}[1]{>{\raggedleft\let\newline\\arraybackslash\hspace{0pt}}m{#1}}

\begin{document}

\begin{titlepage}
\title{Arbitrage and rents in European long-term transmission rights}

\author[]{Clemens Stiewe\thanks{Corresponding author: clemens.stiewe@gmail.com}}
\affil[]{\small \textit{Centre for Sustainability, Hertie School, Berlin}}
\date{July 30, 2026}
\maketitle
\begin{abstract}
Long-term transmission rights (LTTRs) are designed to support hedging in interconnected European electricity markets. LTTR auction prices have historically fallen short of forward market prices, signaling limited arbitrage. This paper studies the interaction of transmission rights and forward markets. Option pricing theory predicts that LTTR holders take short forward positions in importing markets and long forward positions in exporting markets to lock in arbitrage profits. Empirically, I find a corresponding price effect in the German electricity forward market immediately after LTTR auctions, using panel regression on EEX futures contracts traded between 2018 and 2025. This shows that LTTR holders can achieve systematic rents, indicating an inefficient regulatory intervention and a transfer from consumers to LTTR holders. \\
\vspace{0in}
\vspace{0in}

\bigskip
\end{abstract}
\setcounter{page}{0}
\thispagestyle{empty}
\end{titlepage}
\pagebreak \newpage


\section{Introduction} \label{sec:introduction}
Electricity spot market prices are volatile because electricity cannot yet be stored economically. Forward markets therefore provide insurance against spot price volatility for generators (i.e., spot market sellers) and retailers (i.e., spot market buyers). Forward market liquidity is a concern in many European countries, however. Liquidity concentrates on the German forward market, which has substantially higher traded volume and lower bid-ask spreads than neighboring markets \citep{acerMarketmonitoring2025}. Austrian market participants, for example, might thus prefer to hedge on the German forward market. While this offers lower transaction costs, it also comes with ``basis risk'': German futures are settled against the German spot price index, so an Austrian generator selling forward in Germany is exposed not only to Austrian but also to German spot prices. Whenever these spot prices diverge, the hedge becomes incomplete. 
\par
Transmission system operators (TSOs) are therefore required to issue long-term transmission rights (LTTRs) for market borders with substantial liquidity gaps \citep{CommissionRegulationEU2016a}. LTTRs are financial transmission rights (FTRs) and resemble European-style call options on the hourly spot price spread between two neighboring electricity markets during the delivery period. LTTRs are issued for most physical market borders and directions as monthly and annual products.\footnote{Pay-as-clear auctions take place every month and year shortly before the delivery period begins, e.g., from 24 November 2024 until 2 December 2024 for an annual 2025 LTTR. Auction results are published right after the end of the auction window.}
\par
Regulators have acknowledged several design shortcomings that limit the attractiveness of LTTRs as a hedging instrument for generators and retailers, such as the short maximum maturity of one year, the lack of a secondary market, and incomplete financial firmness\footnote{Reduced firmness means that holders face reduced payouts during force majeure events.} \citep{acerMarketmonitoring2025}. Because of these shortcomings, regulators assume that substantial volumes of LTTRs are bought by financial traders rather than by physical hedgers \citep{acerFurtherDevelopmentEU2023,europeancommissionTargetedConsultationRevision2024}. LTTR auction prices fall systematically short of forward market prices \citep{acerMarketmonitoring2024}, which likely reflects that financial traders, who have a lower willingness to pay than physical hedgers, tend to be the marginal, price-setting bidders in LTTR auctions. For inframarginal participants with higher willingness to pay, the underpricing of LTTRs relative to forward markets represents an arbitrage opportunity. Using panel regression on German and Austrian forward contracts traded between 2018 and 2025, I show that market participants exploit this opportunity by selling the forward spread immediately after buying LTTRs, hence locking in profits. Arbitrage across LTTR and forward markets is limited because of transaction costs and inelastic LTTR supply, which explains why mispricing can persist. The rents earned by LTTR holders are financed from congestion income that would otherwise reduce transmission tariffs \citep{RegulationEU20192019}. This study hence provides empirical support for regulators' concerns about a transfer from consumers to LTTR holders \citep{acerFurtherDevelopmentEU2023}. More generally, it studies the interaction of transmission rights with related derivatives markets.
\par
This paper is organized as follows: \autoref{sec:background} reviews the literature on forward market price formation and transmission rights, \autoref{sec:theory} derives the expected effects on forward markets, and \autoref{sec:empirical} discusses the empirical strategy. \autoref{sec:result} and \autoref{sec:discussion} present and discuss the results, while \autoref{sec:conclusion} concludes.

\section{Research background} \label{sec:background}
This section first reviews the relevant literature on electricity forward markets, focusing on how price formation is shaped by the hedging pressures of retailers and generators. Controlling for the effects of physical participants' hedging needs then allows me to empirically isolate the forward market effect of LTTR arbitrage.
\par
Forward prices for most commodities can be explained by storage costs \citep{kaldorSpeculationEconomicStability1939}. Since electricity cannot be stored economically for longer periods, however, theories of storage cannot explain electricity forward prices. The empirical literature on electricity forward prices therefore mostly builds on theories that involve ``hedging pressures'' of buyers and sellers. The concept dates back to \citet{keynesTreatiseMoneyPure1930}, who postulated that whenever there are more sellers than buyers on the forward market, speculators take over the price risk from sellers in exchange for a premium, resulting in ``normal backwardation'', i.e., forward prices that are below expected spot prices. This difference between forward and expected spot prices for the same delivery period has also been referred to as the forward (or forward risk) premium.\footnote{Forward premiums can also emerge when agents are risk-neutral but have market power. Building on the seminal work by \citet{allazCournotCompetitionForward1993}, who introduce a model of Cournot competition on forward markets, \citet{itoSequentialMarketsMarket2016} show that strategic behavior can induce a premium in sequential electricity markets, and empirically confirm this for Spain. \citet{borensteinInefficienciesMarketPower2008} show that besides risk aversion, market power drives forward premiums on the Californian electricity market.}
\par
The equilibrium model of \citet{bessembinderEquilibriumPricingOptimal2002} has been an influential extension of the hedging pressure theory to electricity forward markets. They show that the size and the sign of the forward premium are affected by the skewness and variance of expected spot prices, which are fundamentally determined by the level of electricity demand. When expected spot prices are positively skewed and electricity demand is high, retailers face the risk of spot price spikes, which increases their hedging pressure and induces a positive forward premium. When expected demand and skewness of prices are low, the hedging pressure of generators dominates, and a negative premium emerges. A higher variance of spot prices means that generators face uncertain spot revenues, which increases their hedging pressure and leads to a negative forward premium.
\par
Electricity forward prices, and the forward premium, are also positively linked to the fuel prices that electricity generators have to pay. Gas, coal, and carbon price risks therefore affect not only the spot, but also the forward market price of electricity \citep{redlDeterminantsPremiumForward2011, bunnForwardPremiumElectricity2013}. Higher market shares of wind, solar, and hydro energy, on the other hand, are associated with lower forward premiums \citep{huismanPricingForwardContracts2021,carneroExAnteEx2025}.
\par
Because generators tend to hedge price risk at longer horizons than retailers, their net hedging pressure varies with time to maturity \citep{benthPricingForwardContracts2008}. This notion is also crucial for \citet{fletenOvernightRiskPremium2015}, who document an overnight risk premium in Nordic and German/Austrian electricity futures that is positive prior to the front period of a forward contract, when generators' hedging pressure dominates, and smaller (or even negative) after retailers enter the forward market in the front period once they know how much to sell to end consumers. \citet{penaMarketMakersLiquidity2022} argue that the forward premium in \citet{fletenOvernightRiskPremium2015} is mostly a liquidity premium that decreases in size as the number of market makers, who provide liquidity to physical participants, increases towards maturity. I build on the empirical framework of \citeauthor{fletenOvernightRiskPremium2015} as it allows me to control for hedging pressure effects, isolating the effect of LTTR arbitrage after auctions. An effect of LTTR auctions on forward prices is suggested by \citet{hirthCrossborderForwardMarkets2024} in a consultancy report commissioned by German TSOs, but has not been established empirically in the literature.
\par 
The literature on forward market price formation hence provides the foundation for my identification strategy. A second relevant strand examines the design of FTRs, focusing on US electricity markets. This literature provides empirical evidence that US markets for transmission rights involve large transfers from consumers to financial traders but offers different interpretations.
\par
\citet{opgrandPriceFormationAuctions2022} and \citet{dengInherentInefficiencySimultaneously2010} find that FTR auction proceeds systematically fall short of their expected ex post value. \citeauthor{opgrandPriceFormationAuctions2022} show that physical hedgers are willing to pay a positive risk premium in FTR auctions, while financial traders require a negative trading premium for holding FTRs. This trading premium may contain transaction costs, such as collateral requirements, and a risk premium. FTR underpricing is therefore attributed to a larger share of speculative buyers in the auction. \citet{dengInherentInefficiencySimultaneously2010}, on the other hand, attribute FTR underpricing to an inefficient auction mechanism. Longer-term FTRs, which could be used by physical participants to support project finance, are suggested as a remedy \citep{risangerCongestionRiskTransmission2024}. \citet{petropoulosLongtermTransmissionRights2020} show that FTRs increase dynamic efficiency by reducing premature investment by incumbent players, but only if a secondary market for FTRs exists.
\citet{leslieWhoBenefitsRatepayerfunded2021} empirically studies firm-level positions of retailers, generators, and financial traders in transmission congestion contracts (TCCs), the FTRs of New York's NYISO market, and finds that retailers participate primarily for hedging, while generators and traders earn profits from their TCC positions. This suggests a transfer of wealth from network users to firms. However, \citeauthor{leslieWhoBenefitsRatepayerfunded2021} cautions that the net welfare effect of trader participation in TCC auctions on network users can be ambiguous: trader participation improves price discovery and liquidity, which may ultimately benefit network users through more efficient retail contracts or generation and transmission planning. The participation of financial players generally improves market efficiency, for example through arbitrage \citep{jha2013testing,mercadal2022dynamic}. \citet{birge2018limits} show, however, that financial participation needs to be accompanied by adequate market design: when the regulator imposed higher transaction costs on financial participants in the MISO market, thus limiting their arbitrage profits, financial players started to engage in the profitable manipulation of FTR value instead. 
\par
Inefficiencies of transmission rights on single European borders have been documented \citep{mcinerneyValuationAnomaliesInterconnector2013,kristiansenAssessmentDanishGerman2007}, but the harmonized European LTTRs introduced in 2016 have received little attention in the literature. Regulators have documented underpricing of LTTRs, attributing it primarily to short maturities, infrequent auctions and the lack of a secondary market, as well as incomplete financial firmness \citep{acerMarketmonitoring2024, beatoLongTermInterconnection2021}. A survey of market participants indicates that some of them prefer spread futures over LTTRs for hedging basis risk \citep{europeancommissionTargetedConsultationRevision2024}. \citet{acerMarketmonitoring2024} computes the difference between cleared LTTR auction prices and contemporaneous EEX baseload spread futures prices for the same delivery period, and finds that on most European borders, the cross-border spread is priced lower in LTTR auctions than on EEX forward markets. A systematically higher value for obligations (EEX futures) than for options (LTTRs) on the same underlying is counterintuitive because spreads are usually volatile and can change sign from hour to hour, which should make an option more valuable than an obligation \citep{hullOptionsFuturesOther2021}. Reduced demand from physical hedgers, e.g., because of the non-firmness of LTTRs, and a lower willingness to pay of financial traders, who purchase these transmission rights instead, might explain LTTR undervaluation. In the following section, I show that LTTR underpricing can be profitably exploited through arbitrage and derive the expected effects of that strategy on forward markets.
\par
This paper contributes to the literature by documenting the interaction of transmission rights with related derivatives markets. It provides a first empirical assessment of the efficiency of European LTTRs and extends the empirical literature on forward market price formation with an estimate of the effect of transmission rights auctions.

\section{Expected forward market effects} \label{sec:theory}
Market participants holding LTTRs can exploit mispricing of spreads across LTTR and forward markets by selling the baseload forward spread immediately after acquiring LTTRs. In this section, I derive the expected effect of these forward positions.
\par
The payoff of an $A{\to}B$ LTTR is defined as the monthly or annual sum of positive hourly $B{-}A$ price spreads (i.e., $\sum_{h=1}^{H} \max \bigl( P_h^{B} - P_h^{A} ,\,0 \bigr )$ for annual ($H=8760$) or monthly ($H=730$) LTTRs). Net profits $\Pi$ at delivery $T$ in Euro/MWh from the combined long LTTR and short forward position are given by: 
\begin{equation} \label{eq:arbitrage}
\Pi_T = \underbrace{\frac{1}{H}\sum_{h=1}^{H} \max\bigl(P_h^{B} - P_h^{A},\, 0\bigr)}_{\text{LTTR payoff}}
+ \underbrace{\bigl(F_{t,T} - \frac{1}{H}\sum_{h=1}^{H} \bigl(P_h^{B} - P_h^{A}\bigr)\bigr)}_{\text{forward payoff}}
- L_{t,T} - \tau_t
\end{equation}
\noindent
where $\frac{1}{H}\sum_{h=1}^{H} \bigl(P_h^{B} - P_h^{A}\bigr) \equiv S_T$ is the average cross-border spot price spread during the delivery period, $F_{t,T}$ is the forward spread price on LTTR auction day $t$ for delivery at $T$, $L_{t,T}$ is the auction price paid at time $t$ for delivery at $T$, and $\tau_t$ are the transaction costs of arbitrage priced into the auction bid at time $t$ as part of a (negative) trading premium. The LTTR payoff is expressed per MWh, i.e., divided by $H$.
\par
Suppose a trader participates in the 2025 LTTR auction and bids today's $B{-}A$ Cal-2025 forward spread of 17 Euro/MWh minus a trading premium of 1 Euro/MWh (i.e., 16 Euro/MWh) for one megawatt (MW) of $A{\to}B$ LTTR. Assume for simplicity that the trading premium contains only the transaction costs of arbitrage. On the next day, the auction clears at 15 Euro/MWh because the marginal bidder has priced in transaction costs of 2 Euro/MWh. The $B{-}A$ forward spread still trades at 17 Euro/MWh. To capture the gap between the clearing price and the forward spread price, she sells the $B{-}A$ spread forward (i.e., selling one MW forward in $B$ and buying one MW forward in $A$) and holds both positions until delivery. If the realized spread is positive during all hours of 2025, the LTTR payoff in \autoref{eq:arbitrage} reduces to $S_T$, and $\Pi_T = S_T + (F_{t,T} - S_T) - L_{t,T} - \tau_t=F_{t,T}-L_{t,T}-\tau_t = 17-15-1=1$, leaving the trader with a 1 Euro/MWh profit. If the spread turns negative in some hours, the LTTR does not pay off in those hours while $S_T$ falls, such that the short forward payoff increases. The trader's payoff then equals 1 Euro/MWh plus an uncertain upside from negative-spread hours.
\par
This strategy is only profitable with LTTRs that export from low-price to high-price zones: assuming that the $B{\to}A$ LTTR clears at a price of zero, selling the $A{-}B$ spread forward at -17 Euro/MWh effectively means paying a large fixed amount ($-17-0-1=-18$) for a small optionality (because the $A{-}B$ spread may only turn positive during a few hours).
\par
Buying the forward spread low and selling it high as an LTTR is not possible: without a secondary market for LTTRs, TSOs are the only sellers of LTTRs. Market participants can thus only lock in arbitrage profits if they can buy the cross-border spread low (in LTTR auctions) and sell it higher (on forward markets). Arbitrage with LTTRs from low-price to high-price markets implies selling forward in the high-price zone and buying forward in the low-price zone. Assuming that traders hold equal volumes of LTTRs in both directions of each border, their net forward positions should be short in high-price, importing and long in low-price, exporting markets.
\par
Transaction costs $\tau_t$ include funding liquidity costs and market liquidity costs \citep{brunnermeierMarketLiquidityFunding2009}. Spread volatility raises the risk of large and frequent margin calls, forcing the trader to post variation margin. Funding liquidity costs are smaller if spreads are stable and the trader can reliably hold the position to delivery. The directional effect on forward markets should hence be strongest when cross-border spreads are stable. Market liquidity costs are determined by bid-ask spreads, which are larger in illiquid markets.
\par
\begin{figure}[htbp]
  \centering
  \includegraphics[width=\linewidth]{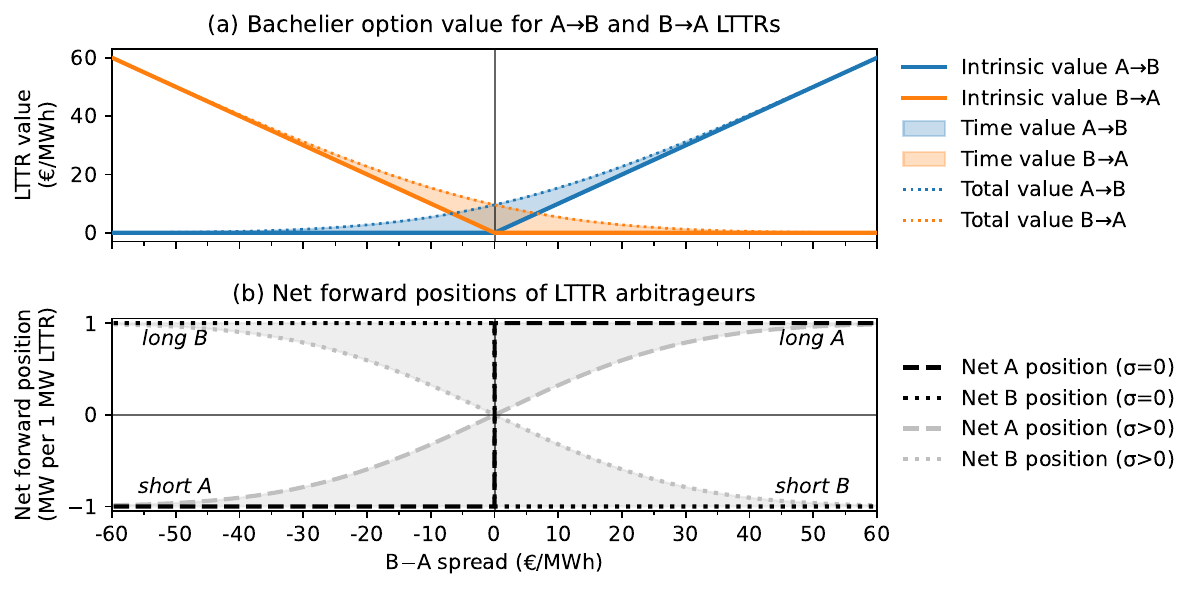}%
  \caption{Bachelier option values and net forward positions vs.\ B--A spread. Panel (a) shows how LTTR option value changes with the cross-border spread. The corresponding net forward positions of arbitrageurs are shown in (b). Spread volatility increases option value but reduces the size of forward market positions. Net positions are short in importing and long in exporting markets.}
  \label{fig:lttr_values_and_net_B}
\end{figure}

Using the Bachelier option pricing model\footnote{The undiscounted Bachelier price of a call option on an underlying with forward price $F_0$, option strike price $K$, time-to-maturity $T$ and volatility of the underlying $\sigma_N$ is given as $C_N{(K)} = (F_0 - K)\,N\!\left( d_N \right) + \sigma_N \sqrt{T}\,n\!\left( d_N \right)$, where $d_N = \frac{F_0 - K}{\sigma_N \sqrt{T}}$, $N(z)$ is the cumulative distribution function and $n(z)$ is the probability density function of the standard normal distribution \citep{choiBlackScholesUsers2022}.} \citep{bachelierTheorieSpeculation1900}, \autoref{fig:lttr_values_and_net_B} illustrates how the expected aggregate effect of this strategy depends on the level and the volatility of the cross-border spot price spread. Panel (a) plots the Bachelier option value for hypothetical $A{\to}B$ and $B{\to}A$ LTTRs across a range of $B{-}A$ spreads.
\par
Panel (b) shows the net forward positions in markets $A$ and $B$. In the unrealistic case of a perfectly stable spread ($\sigma=0$), the LTTR only has intrinsic value and arbitrage transaction costs are small. This makes it relatively more attractive for traders to lock in profits via arbitrage. Volatility ($\sigma>0$) increases both option value and transaction costs, making it relatively more attractive to keep the LTTR option on the books instead. For a given positive volatility, a larger average positive spread means that negative-spread hours are less likely. This reduces the uncertain upside from the combined LTTR/forward position and hence the volatility of the trader's portfolio. 
\par
Aggregate forward positions, which reflect the difference in $A{\to}B$ and $B{\to}A$ option deltas from panel (a), and thus the effect on forward market prices, should therefore be strongest when spreads are large and stable.

\section{Empirical strategy} \label{sec:empirical}
Empirically testing these hypotheses requires a model of forward prices that is granular enough to capture the instantaneous impact of LTTR arbitrage while controlling for the effects of physical hedging on forward prices. I build on the model of \citet{fletenOvernightRiskPremium2015} and use the daily log returns of electricity futures, i.e., the first differences of the natural logarithm of daily settlement prices for the respective futures contract, as the dependent variable. In what follows I introduce the independent variables, explain the identification strategy and the model specification, and describe the data.

\subsection{Measuring the effect of LTTR auctions} \label{sec:measurement}
\begin{figure}[htbp]
  \centering
  \includegraphics[width=\linewidth]{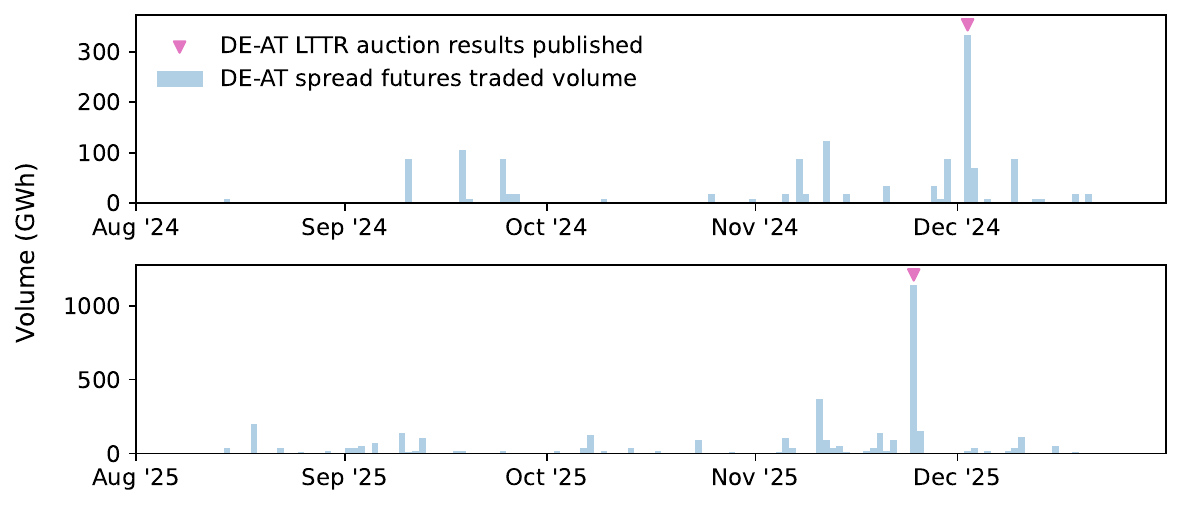}%
  \caption{DE-AT spread futures volumes. Daily spread futures volumes traded at EEX exhibit a spike on the day when LTTR auction results for the DE-AT and AT-DE LTTRs are published, suggesting that market participants open spread futures positions right after auctions.}
  \label{fig:spread_futures}
\end{figure}

\begin{figure}[htbp]
  \centering
  \includegraphics[width=\textwidth]{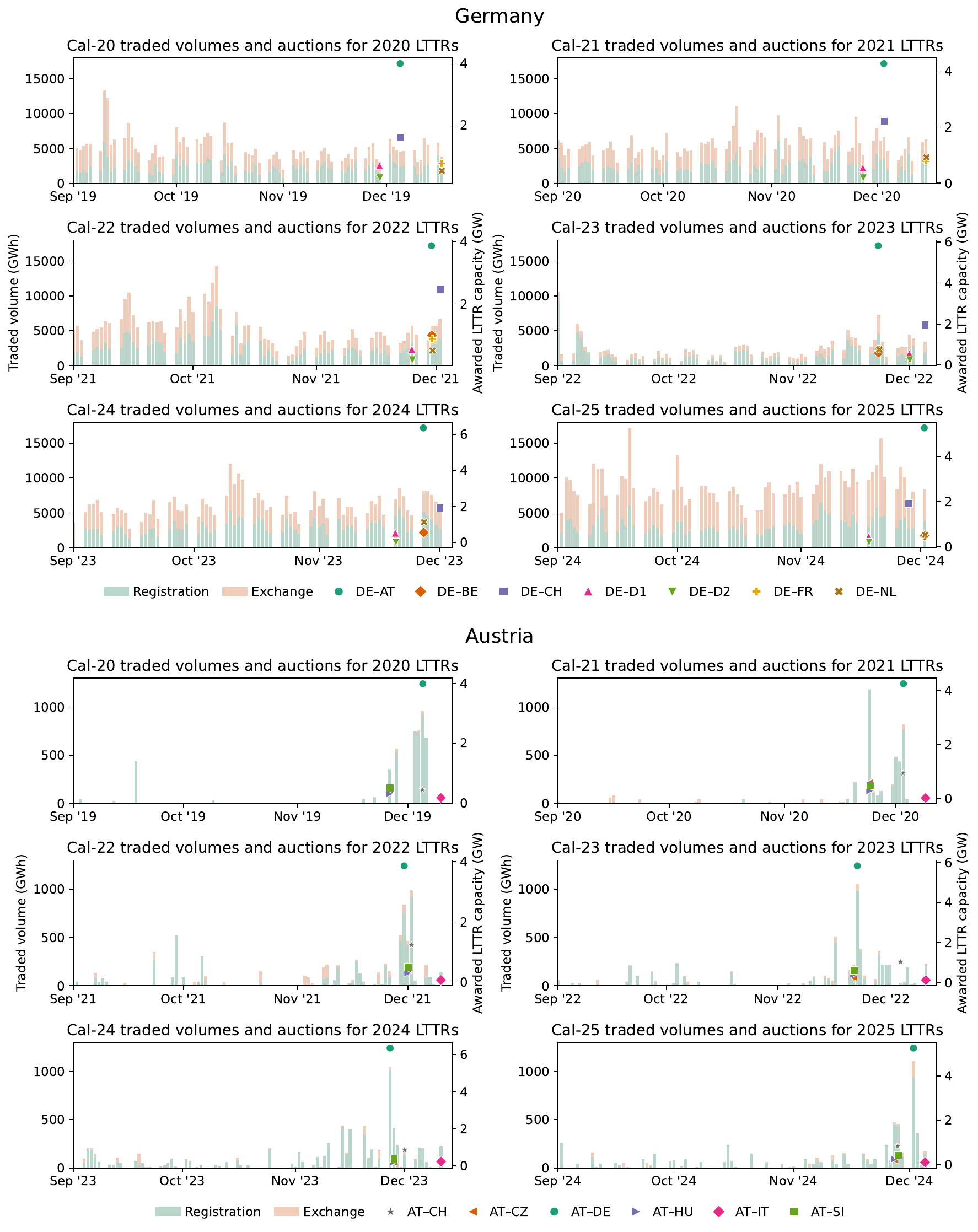}%
  \caption{Front-year futures traded volumes and annual LTTR auctions for Germany and Austria. Total traded futures volume comprises exchange-traded volumes and cleared bilateral trades registered at EEX. Total awarded LTTR capacity per border (summed up across directions) is shown on the secondary axis.}
  \label{fig:volumes_DE_AT}
\end{figure}

\autoref{sec:theory} suggests that new forward positions should be opened after the publication of LTTR auction results. Auctions are usually clustered around the beginning of December for annual and around the third week of a month for monthly LTTRs. \autoref{fig:spread_futures} shows daily traded volumes for annual German-Austrian spread futures, which are equivalent to simultaneous buy and sell orders for two outright futures, one in Germany and one in Austria. The publication of German-Austrian LTTR auction results coincides with clear spikes in traded volumes of spread futures with the same delivery period, suggesting that some LTTR holders engage in arbitrage immediately after acquiring transmission rights. These trades around auctions, not only for the German-Austrian border but also for others, are visible in total outright traded volumes too, especially for Austria (see \autoref{fig:volumes_DE_AT}).
\par
I construct separate panels of annual and monthly futures contracts $i$ traded at daily resolution $t$ to test the effect of LTTR arbitrage on the returns of futures contracts with the same delivery period. This means, for example, testing whether overnight returns of the Cal-2025 base future are driven by auction results for an annual LTTR with delivery in 2025.
\par
To capture the instantaneous effect of LTTR auctions, I construct $Q_{i,t}$, which reflects awarded LTTR capacity per trading day, normalized by total awarded LTTR capacity for the respective delivery period. $Q_{i,t}$ therefore represents the normalized daily sum of bilateral LTTR capacities shown in \autoref{fig:volumes_DE_AT} and is zero on non-auction days.
\par
While the sign of the effect of arbitrage depends on which LTTR direction is in the money (i.e., on the sign of the cross-border spread), the size of the effect should increase with auctioned capacities and spread levels and should decrease with spread volatility (see \autoref{fig:lttr_values_and_net_B}). \par
\clearpage
To approximate expected spread levels, I calculate the difference between observed auction prices of importing and exporting LTTRs per auction day. To reflect that LTTR capacities differ by direction for some borders, I weight importing and exporting auction prices by the capacities awarded in the respective direction. That is, for each futures contract $i$ on market $A$ with neighbors $j=1,\dots,J$\footnote{For Germany, $J$ includes Danish bidding zones DK1 and DK2, the Netherlands, Belgium, France, Switzerland, Austria, as well as the joint Polish-Czech LTTR profile. For Austria, $J$ includes Switzerland, Italy, Slovenia, Hungary, Czechia, and Germany.} and day $t$ I define:

\begin{equation}
\begin{aligned}
{LTTR}^{A}_{i,t}
= \frac{\sum_{j=1}^J C_{i,t}^{j \to A}\, P_{i,t}^{j \to A}}{\sum_{j=1}^J C_{i,t}^{j \to A}}
- \frac{\sum_{j=1}^J C_{i,t}^{A \to j}\, P_{i,t}^{A \to j}}{\sum_{j=1}^J C_{i,t}^{A \to j}} 
\end{aligned}
\end{equation}
where $P^{A \to j}_{i,t}$ and $P^{j \to A}_{i,t}$ denote the LTTR auction prices in direction $A {\to}j$ and $j{\to}A$, and $C^{A \to j}_{i,t}, C^{j \to A}_{i,t}$ are the corresponding awarded capacities. ${LTTR}^{A}_{i,t}$ is hence denominated in Euro/MWh. A positive value of ${LTTR}^{A}_{i,t}$ reflects that the importing ($j{\to}A$) LTTR has the higher expected payoff, such that the expected effect of ${LTTR}^{A}_{i,t}$ on overnight returns is negative: the more positive the LTTR price spread (i.e., the higher the expected spot price in $A$, relative to its neighbors), the larger the short forward positions of LTTR arbitrageurs in $A$. Increased supply of forward contracts should reduce equilibrium forward prices on auction days, resulting in negative overnight returns.
\par
To approximate spread volatility, I define $\sigma^{A}_{i,t}$ as the standard deviation of daily changes in the $A{-}j$ spot price spread over the last two years before the respective LTTR auction. On days with auctions on multiple borders, I calculate a capacity-weighted average volatility across all neighbors. Similar to ${LTTR}^{A}_{i,t}$, $\sigma^{A}_{i,t}$ captures the spread volatility per auctioned border and is zero on non-auction days.

\subsection{Identification} \label{sec:identification}
A potential threat to identification is that daily futures returns may themselves influence LTTR auction results. The auction window for annual LTTRs usually spans seven days, while the window for monthly products is usually between two and four days. Auction results are published around 20 minutes after the auction window is closed at 2-4pm CET. The window for settlement price formation at EEX, on the other hand, is 5:05-5:15pm CET since 2022 (it was 4:20pm-4:30pm CET prior to 2022) \citep{ClearingCircular422022}. Even if LTTR auction bids are submitted at the very end of the auction window, the LTTR spread is determined prior to futures prices, which qualifies the LTTR spread as contemporaneously exogenous and allows for a causal interpretation of Ordinary Least Squares (OLS) estimates \citep{wooldridgeEconometricAnalysisCross2010}. 
\par
Past futures returns may affect current futures returns, for example through the bid-ask spread \citep{rollSimpleImplicitMeasure1984}. At the same time, past futures returns may affect the fair value of LTTRs and thus the current LTTR spread. Conditioning on lagged returns should block these causal paths.
\par
Spot market conditions determine both forward prices and the fair value of LTTRs. Hence, it is crucial to control for common fundamentals and systematic risk. Following the literature, I therefore include fuel and equity market returns, renewable generation, and the variance and skewness of spot prices. Moreover, specifying the dependent variable as a first difference reduces persistence and spurious correlation with the LTTR spread, which is a level variable \citep{hamiltonTimeSeriesAnalysis1994}. 
\par
Another identifying assumption is that LTTR auctions only affect returns on the days when LTTR auction results are published. This assumption is tested with an event study around auction days. Even with flat pre- and post-auction trends, however, it cannot be ruled out that LTTR auction days coincide with unobserved events that affect forward prices. I therefore use the event study design to conduct a placebo test, which does not use the front-year futures, as in the baseline model, but the second year-ahead returns as the dependent variable. This means, for example, testing whether the auction for the 2025 LTTR has an effect on Cal-2026 futures returns. Both Cal-2025 and Cal-2026 futures are traded on LTTR auction days, but only the Cal-2025 hedges the 2025 LTTR. Hence, Cal-2026 returns should be unaffected by the auctions for the 2025 LTTR.

\subsection{Model specification} \label{sec:model}
I estimate the following panel model for the log returns $r_{i,t}$ of annual or monthly contracts $i$ at daily resolution $t$:
\begin{equation}
\begin{aligned}
r_{i,t} &= \beta_{0}
+ \beta_{1}\, Q_{i,t}\cdot LTTR_{i,t}
+ \beta_{2}\, Q_{i,t}\cdot LTTR_{i,t}\cdot \sigma_{i,t} \\
&\quad + \beta_{3}\, r_{i,t-1}
+ \beta_{4}\, Q_{i,t}
+ \beta_{5}\, LTTR_{i,t} 
+ \beta_{6} \sigma_{i,t} \\
&\quad + \beta_{7}\, Gas_{t}
+ \beta_{8}\, Coal_{t}
+ \beta_{9}\, Carbon_{t}
+ \beta_{10}\, Stock_{t} \\
&\quad + \beta_{11}\, Wind_{t}
+ \beta_{12}\, Solar_{t}
+ \beta_{13}\, Reservoir_{t} \\
&\quad + \beta_{14}\, Var_{t}
+ \beta_{15}\, Skew_{t} \\
&\quad + \alpha_{i} + \gamma_{m} + \gamma_{dow} + \gamma_{hol} + \gamma_{front} + \gamma_{2022} + \epsilon_{i,t}
\end{aligned}
\end{equation}
where $Q_{i,t} \cdot {LTTR}_{i,t}$ is the variable of interest: it captures how the impact on forward returns scales with the volume of auctioned capacity and the LTTR spread. A potentially attenuated effect on forward positions at higher spread volatility is captured by $Q_{i,t} \cdot {LTTR}_{i,t} \cdot \sigma_{i,t}$.
\par
$Skew_{t}$ and $Var_{t}$ are skewness and variance of spot prices calculated over a 365-day historical rolling window. I control for the effect of fuel prices by including the logarithmic returns of TTF gas ($Gas_{t}$), API2 CIF ARA coal ($Coal_{t}$), and EU ETS ($Carbon_{t}$) prices. $Wind_{t}$ and $Solar_{t}$ are the first differences of daily wind and solar generation, while $Reservoir_{t}$ controls for the deviations of daily hydro reservoir filling ratios from historical averages. I control for equity market conditions by including the logarithmic returns of national stock market indices ($Stock_{t}$). $\alpha_{i}$, $\gamma_{m}$, $\gamma_{dow}$, $\gamma_{front}$, and $\gamma_{hol}$ capture contract, month, day-of-the-week, front-period and holiday fixed effects. The 2022 energy crisis was characterized by high price volatility and substantially reduced forward market liquidity \citep{krogerContractsCrisisWar2025}. To estimate the effect of LTTR arbitrage during non-crisis conditions, I control for the extraordinary level and slope effects of the 2022 energy crisis through $\gamma_{2022}$, which contains a 2022 dummy and an interaction of that dummy with $Q_{i,t} \cdot {LTTR}_{i,t}$. The first lag of returns $r_{i,t-1}$ is included to account for autocorrelation in daily returns and the effect of past forward prices on current LTTR spreads. I compute Driscoll–Kraay standard errors to make inference robust to heteroskedasticity, autocorrelation, and cross-sectional dependence. This accounts for persistent common shocks that affect multiple contracts.

\subsection{Data} \label{sec:data}
I estimate the model separately for the German and Austrian forward markets. The German market is the largest and most liquid forward market in Europe, while the neighboring Austrian market is much less liquid (see \autoref{fig:volumes_DE_AT}). Yet, relative to total electricity consumption, Austria is more interconnected than Germany \citep{stieweCrossborderCannibalizationSpillover2025}. Austrian LTTR capacities are therefore relatively larger than German capacities, suggesting, other things equal, a stronger effect of LTTR arbitrage in Austria. I use daily trading data from 2018 to 2025 on annual and monthly base and peak futures contracts traded at EEX. The final sample consists of 26 annual and 186 monthly base and peak contracts each for the German and Austrian markets. Wind, solar, and hydro reservoir data are obtained from \citet{entso-eENTSOETransparencyPlatform2026}, while LTTR auction results are published by \citet{jointallocationofficeJointAllocationOffice2026}. Natural gas prices are retrieved from \citet{investing.comDutchTTFNatural2026}, coal prices from \citet{investing.comCoalAPI2CIF2026}, carbon prices from \citet{AllowancePriceExplorer}, and stock indices from \citet{investing.comDAX} and \citet{investing.comATX}.
\par
I thus estimate four separate regressions: returns of monthly (annual) futures on monthly (annual) LTTR spreads for Germany and Austria. In these regressions, both base and peak futures contracts are included. Reservoir filling ratios are only controlled for in the Austrian regressions as data are not available for Germany and the share of hydro reservoirs in electricity generation is small. 
\autoref{tab:descriptives} shows that annual LTTRs importing into Germany have on average been priced slightly higher than LTTRs exporting from Germany, while the opposite holds for annual LTTRs with a leg in Austria. The wide range of LTTR spreads for both countries and durations reflects that trade patterns change over time. This is also evident in \autoref{fig:lttr_spread}, which plots the daily spreads in monthly and annual LTTR auction prices for Germany and Austria. The econometric analysis aims to exploit this variation.
\begin{table}[htbp]
\centering
\caption{Descriptive statistics}
\label{tab:descriptives}
\begin{tabular}{lrrrrrr}
\hline
Variable & Mean & Median & Min & Max & Std & Skew \\
\hline
\noalign{\vskip 0.5ex}
\textbf{Germany} & & & & & & \\
${LTTR}_{i,t}$ (annual, EUR/MWh) & 0.6 & -2.2 & -22.2 & 33.3 & 10.5 & 1.0 \\
${LTTR}_{i,t}$ (monthly, EUR/MWh) & -3.2 & -1.4 & -90.2 & 52.7 & 13.5 & -1.8 \\
$Q_{i,t}$ (annual, GW) & 3.4 & 1.1 & 0.4 & 15.5 & 4.5 & 1.6 \\
$Q_{i,t}$ (monthly, GW) & 3.8 & 1.8 & 0.1 & 21.8 & 4.3 & 1.3 \\
$\sigma_{i,t}$ & 16.5 & 13.8 & 5.9 & 39.7 & 9.8 & 1.1 \\
$Var_{t}$ & 3381.0 & 919.8 & 107.0 & 20137.1 & 5581.7 & 1.9 \\
$Skew_{t}$ & 0.1 & -0.2 & -2.4 & 3.0 & 1.2 & 0.1 \\
$Stock_{t}$ & 15390 & 14494 & 8442 & 24611 & 3687 & 1 \\
$Wind_{t}$ (GW) & 11.8 & 9.4 & 0.3 & 44.3 & 8.7 & 1.1 \\
$Solar_{t}$ (GW) & 6.1 & 5.7 & 0.2 & 20.0 & 4.2 & 0.5 \\
\noalign{\vskip 0.5ex}
\textbf{Austria} & & & & & & \\
${LTTR}_{i,t}$ (annual, EUR/MWh) & -2.1 & -3.0 & -22.2 & 17.5 & 7.2 & 0.1 \\
${LTTR}_{i,t}$ (monthly, EUR/MWh) & -4.0 & -3.0 & -47.2 & 47.0 & 10.4 & -0.1 \\
$Q_{i,t}$ (annual, GW) & 3.3 & 1.5 & 0.1 & 12.2 & 4.3 & 1.2 \\
$Q_{i,t}$ (monthly, GW) & 2.7 & 0.7 & 0.0 & 20.3 & 4.0 & 1.5 \\
$\sigma_{i,t}$ & 14.4 & 12.8 & 5.7 & 30.3 & 7.6 & 0.7 \\
$Var_{t}$ & 3462.2 & 810.9 & 59.9 & 21891.0 & 5799.0 & 1.9 \\
$Skew_{t}$ & 0.3 & 0.1 & -1.7 & 3.1 & 1.0 & -0.0 \\
$Stock_{t}$ & 3327 & 3262 & 1631 & 5326 & 593 & 0 \\
$Wind_{t}$ (GW) & 0.9 & 0.7 & 0.0 & 3.3 & 0.7 & 1.0 \\
$Solar_{t}$ (GW) & 0.3 & 0.1 & 0.0 & 1.5 & 0.3 & 2.1 \\
$Reservoir_{t}$ (GWh) & 74.4 & 65.5 & -277.3 & 508.8 & 169.2 & 0.2 \\
\noalign{\vskip 0.5ex}
\textbf{Common controls} & & & & & & \\
$Carbon_{t}$ (EUR/t) & 54 & 62 & 8 & 98 & 26 & 0 \\
$Gas_{t}$ (EUR/MWh) & 43 & 31 & 4 & 339 & 43 & 3 \\
$Coal_{t}$ (EUR/t) & 109 & 89 & 35 & 403 & 75 & 2 \\
\hline
\end{tabular}
\begin{flushleft}\footnotesize
\vspace{-0.25cm}
\hspace{0.4cm} \textit{Notes}: This table reports summary statistics of untransformed variables.
\end{flushleft}
\end{table}

\section{Results} \label{sec:result}
In this section, I summarize the regression results for Germany and Austria, starting with the models for annual LTTR auctions.
\subsection{Annual LTTR auctions} \label{sec:annual_results}
I find a significant effect of LTTR auction results on forward market returns for Germany in line with the predictions derived in \autoref{sec:theory}. More specifically, overnight returns of annual German futures are more negative when the auction-specific LTTR price spread is more positive, reflecting that importing LTTRs are priced higher than exporting LTTRs. Arbitrage hence reduces prices in importing markets and increases prices in exporting markets. This reduces forward spreads and therefore the gap between forward spreads and LTTR prices. The size of this effect increases with auctioned capacity. It shrinks towards zero at high volatility ($\beta_2$ is positive and significant), which also corroborates my hypothesis. When importing and exporting LTTRs are priced similarly, no clear directional effect on forward markets is expected. Indeed, the main effect of $Q_{i,t}$ (i.e., the auction effect when ${LTTR}_{i,t}$ is zero) is insignificant. 
\par
\begin{figure}[htbp]
  \centering
  \includegraphics[width=\textwidth]{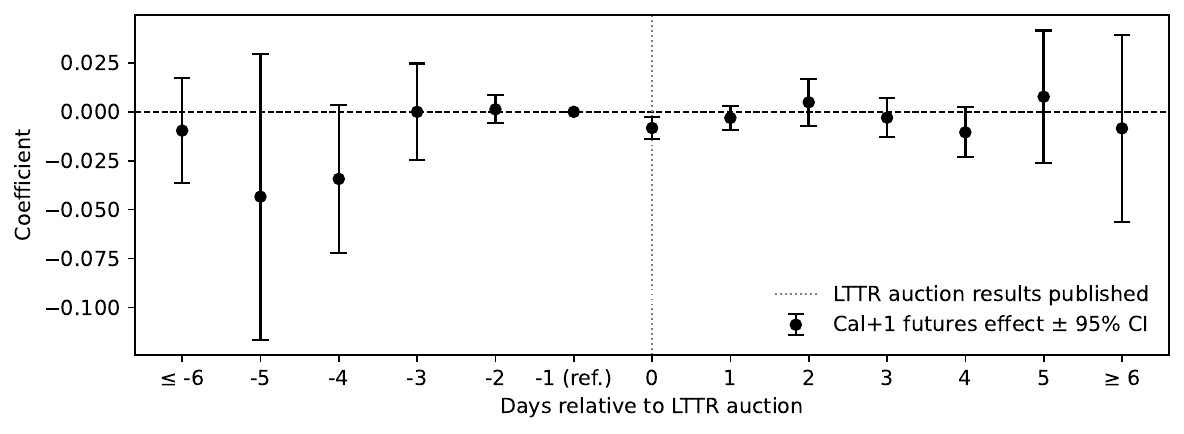}%
  \caption{Effect of LTTR spread around annual LTTR auction days on annual German front-year futures. This event study shows how the effect of $Q_{i,t} \cdot {LTTR}_{i,t}$ evolves around the publication of auction results. Returns are flat before and after auctions, suggesting an immediate and non-persistent price impact of LTTR arbitrage.}
    \label{fig:event_study}
\end{figure}

To assess if LTTR auction results affect returns before and after auctions, I run an event study robustness check on annual German futures (see \autoref{fig:event_study}). The contemporaneous coefficient on $Q_{i,t} \cdot {LTTR}_{i,t}$ is significant while leads and lags are not, which shows that arbitrage has an immediate and non-persistent effect on forward markets after auction results are published. This indicates that the volume spikes in \autoref{fig:spread_futures} reflect market participants realizing profits shortly after the auctions. The event-study placebo test yields an insignificant contemporaneous effect of LTTR auctions on second year-ahead futures returns (see \autoref{fig:event_study_nextcal}), suggesting that the auction-day effects on front-year futures returns are driven by LTTR arbitrage. Apart from a significant effect of lag 5, pre- and post-trends are as flat as front-year returns.
\par
The effect of LTTR auctions on Austrian futures returns, on the other hand, is insignificant. This is surprising given that the spikes in traded Austrian futures volume around LTTR auctions are more pronounced than for Germany (see \autoref{fig:volumes_DE_AT}). However, unlike on the German forward market, most of these traded Austrian volumes are cleared bilateral trades. These trades are not directly taken into account for settlement price formation by EEX. They would affect settlement prices indirectly if, for example, the net short positions of arbitrageurs were offset bilaterally by buyers that would otherwise have traded at the exchange. However, Austrian exchange-traded volumes are generally very small, which suggests that traded volumes on auction days would have been similarly small without LTTR arbitrage. Hence, the indirect effect of cleared bilateral trades on Austrian futures settlement prices is likely negligible, which may explain the insignificant effect.
\par
The control variables enter with signs and magnitudes largely consistent with the literature on forward price formation, suggesting that observable fundamentals are captured well. While gas, carbon, and coal returns exhibit strong and positive effects on futures returns, renewable generation (wind, solar, and reservoir hydro filling ratios) has a negative, yet insignificant effect on futures returns. I find mixed support for the hypotheses of \citet{bessembinderEquilibriumPricingOptimal2002} as only annual German overnight returns show a positive effect of spot price skewness and a negative effect of spot price variance. Full regression results are given in \autoref{tab:regression}.

\par

\begin{table}[htbp]
\centering
\caption{Regression results}
\label{tab:regression}
\begin{tabular}{l cccc}
\hline
 & \multicolumn{2}{c}{\rule{0pt}{1.1em}\textbf{Germany}} & \multicolumn{2}{c}{\rule{0pt}{1.1em}\textbf{Austria}} \\
 & Annual & Monthly & Annual & Monthly \\
\hline
Intercept & -0.0021* & -0.0001 & -0.0021* & 0.0004 \\
 & (0.0012) & (0.0023) & (0.0011) & (0.0023) \\
$Q_{i,t} \cdot {LTTR}_{i,t}$ & -0.0094*** & -0.0017 & 0.0006 & 0.0089 \\
 & (0.0031) & (0.0103) & (0.0040) & (0.0088) \\
$Q_{i,t} \cdot {LTTR}_{i,t} \cdot \sigma_{i,t}$ & 0.0005*** & 0.0005 & -0.0000 & -0.0008 \\
 & (0.0001) & (0.0006) & (0.0002) & (0.0006) \\
$Q_{i,t} \cdot {LTTR}_{i,t} \cdot \gamma_{2022}$ & -0.0042* & -0.0134* & 0.0010 & 0.0102 \\
 & (0.0023) & (0.0076) & (0.0016) & (0.0072) \\
${LTTR}_{i,t}$ & -0.0000 & 0.0001 & -0.0009* & 0.0000 \\
 & (0.0002) & (0.0001) & (0.0005) & (0.0001) \\
$\sigma_{i,t}$ & -0.0000 & 0.0000 & -0.0008*** & 0.0001 \\
 & (0.0001) & (0.0001) & (0.0003) & (0.0001) \\
$Q_{i,t}$ & -0.0095 & 0.0677* & 0.0178* & 0.0288 \\
 & (0.0100) & (0.0363) & (0.0094) & (0.0294) \\
$Gas_{t}$ & 0.1432*** & 0.4661*** & 0.1371*** & 0.4504*** \\
 & (0.0134) & (0.0437) & (0.0130) & (0.0433) \\
$Coal_{t}$ & 0.0242 & 0.1271*** & 0.0218 & 0.1260*** \\
 & (0.0173) & (0.0345) & (0.0169) & (0.0326) \\
$Carbon_{t}$ & 0.2117*** & 0.2015*** & 0.1968*** & 0.1822*** \\
 & (0.0149) & (0.0476) & (0.0141) & (0.0449) \\
$Wind_{t}$ & -0.0004 & -0.0003 & -0.0001 & -0.0001 \\
 & (0.0003) & (0.0007) & (0.0003) & (0.0005) \\
$Solar_{t}$ & -0.0002 & -0.0004 & -0.0000 & 0.0006 \\
 & (0.0006) & (0.0012) & (0.0005) & (0.0011) \\
$Reservoir_{t}$ &  &  & -0.0000 & -0.0000*** \\
 &  &  & (0.0000) & (0.0000) \\
$Stock_{t}$ & 0.0010 & -0.1366** & 0.0280 & -0.0704 \\
 & (0.0241) & (0.0687) & (0.0221) & (0.0694) \\
$Var_{t}$ & -0.0000** & -0.0000** & -0.0000** & -0.0000** \\
 & (0.0000) & (0.0000) & (0.0000) & (0.0000) \\
$Skew_{t}$ & 0.0005** & -0.0005 & 0.0005 & -0.0012* \\
 & (0.0002) & (0.0007) & (0.0003) & (0.0007) \\
$\gamma_{2022}$ & 0.0027* & -0.0098** & 0.0032* & -0.0086** \\
 & (0.0017) & (0.0038) & (0.0017) & (0.0037) \\
$\gamma_{front}$ & -0.0001 & -0.0017*** & -0.0002 & -0.0017*** \\
 & (0.0005) & (0.0004) & (0.0005) & (0.0004) \\
Month, day-of-week, holiday FE & Yes & Yes & Yes & Yes \\
Contract FE & Yes & Yes & Yes & Yes \\
\hline
N & 24168 & 22701 & 24072 & 22620 \\
$R^2$ & 0.32 & 0.52 & 0.32 & 0.50 \\
F-statistic & 350.73 & 730.06 & 325.76 & 664.77 \\
\hline
\end{tabular}
\begin{flushleft}\footnotesize
\vspace{-0.25cm}
\textit{Notes}: Driscoll--Kraay standard errors in parentheses. * p $<$ 0.10, ** p $<$ 0.05, *** p $<$ 0.01
\end{flushleft}
\end{table}

\subsection{Monthly LTTR auctions} \label{sec:monthly_results}
The monthly regressions yield insignificant effects of LTTR auctions, suggesting that holders of monthly LTTRs do not engage in arbitrage after monthly auctions. This might reflect that uncertainty about realized spreads is lower for monthly products. Holders can better forecast next month's spreads than next year's spreads, which makes holding the LTTR until delivery less risky. At the same time, the market liquidity costs of arbitrage may be higher as monthly futures are less liquid than annual futures. 
\par
Taken together, this should make it relatively more attractive to hold monthly LTTRs until delivery. Market participants may therefore equally realize profits with monthly LTTRs, just without a detectable forward market footprint.

\section{Discussion} \label{sec:discussion}
These results show a systematic effect of LTTR auctions on forward returns for annual products on the liquid German forward market. The sign and size of this effect correspond to the price spread in LTTRs, the auctioned capacity, and the volatility of cross-border spreads, which suggests that this forward market footprint reflects arbitrage across transmission rights and forward markets. Despite the insignificant effect, arbitrage likely takes place on the Austrian forward market too: \autoref{fig:volumes_DE_AT} shows a clear correlation of volume spikes and auctions for annual products. Illiquidity, however, likely mutes the price impact of arbitrage in Austria.
\par
Arbitrage hence narrows but, as shown by \citet{acerMarketmonitoring2024}, does not eliminate the gap between LTTR and spread forward prices. This suggests limits to arbitrage \citep{shleiferLimitsArbitrage1997}. Transaction costs represent a first limit to arbitrage \citep{brunnermeierMarketLiquidityFunding2009}. They are part of the trading premium that financial traders demand for holding transmission rights \citep{opgrandPriceFormationAuctions2022}. Because of low demand from physical hedgers for LTTRs, financial participants can set the price in auctions. The negative trading premiums priced into their marginal bids explain why underpricing emerges. If the marginal bidder's trading premium only contains the transaction costs of arbitrage, \autoref{eq:arbitrage} shows that she would just break even when engaging in arbitrage. For inframarginal participants with higher willingness to pay, e.g., because of lower risk aversion or lower transaction costs, arbitrage is profitable. Because transaction costs are likely non-negligible for most participants, however, not all of them engage in arbitrage, preventing mispricing from being fully corrected.
\par
Inelastic supply of transmission rights is another limit to arbitrage. Market participants cannot buy more than the fixed LTTR capacity, which limits the size and the price impact of their orders on forward markets. Even if forward volumes are large relative to average liquidity levels, the example of Austria illustrates that forward market illiquidity can result in muted price effects.
\par
These findings have clear distributional implications. Who benefits from LTTR auctions can be inferred from the structure of the demand curve for LTTRs, which is populated by physical and financial participants. Assuming that all participants are risk-averse, physical hedgers -- if they participate in the auction at all -- should be willing to pay a positive risk premium to hedge basis risk. Financial participants, who demand a premium for holding LTTRs, should hence be placed below them on the LTTR demand curve. If trading premiums are heterogeneous, the marginal, price-setting bidder is the financial trader with the largest negative trading premium. Awarded participants with less negative (or even positive) premiums therefore capture inframarginal rents. A different pricing rule might also result in rents: in pay-as-bid instead of pay-as-clear auctions, bidders might shade their bids down towards the expected marginal price \citep{hortacsuMechanismChoiceStrategic2010}.
\par
Heterogeneity in trading premiums and the higher willingness to pay of physical participants explain rents when all participants are price takers. Market power on the demand side can increase those rents. \citet{ausubel2014demand} show that a bidder whose demand is large relative to the offered capacity has an incentive to bid below her valuation in order to lower the clearing price on the units she wins. An assessment of competition levels in LTTR auctions is beyond the scope of this study but the fact that LTTR auctions are usually several times oversubscribed suggests that price reductions from withholding demand, and thus the profits from strategic behavior, are limited. While the results suggest that LTTR holders are earning rents, a quantification of these rents is not possible in this study because firm-level data is unavailable for Europe. This is another challenge for future research.
\par
TSOs lose from issuing LTTRs, and they lose more than LTTR holders win: because arbitrage involves transaction costs, holders can only realize part of the gap between the LTTR and forward spread price as a profit. TSO losses, on the other hand, amount to the full gap between the fair option value at the time of the auction and the auction price. This shortfall is not borne by TSOs but is financed from the congestion income pool that would otherwise reduce grid fees. If market participants can realize nearly risk-free profits with LTTRs and TSOs can pass through their losses, this only leaves consumers to bear the cost through grid fees. This study hence confirms the concerns of regulators \citep{acerFurtherDevelopmentEU2023}. 
\par
This paper does not attempt a full welfare analysis of LTTRs. Such an analysis should take into account that consumers benefit from increased forward market liquidity and improved price discovery, which might partly offset the transfer from consumers to LTTR holders. These benefits are most likely small, however, as the LTTR-induced forward volumes are essentially one-off events (see \autoref{fig:volumes_DE_AT} for Austria).

\section{Conclusion} \label{sec:conclusion}
This study documents the interaction of transmission rights with forward markets and assesses the efficiency of European LTTRs. Because LTTRs are largely unattractive as a hedging instrument against basis risk, substantial capacity is acquired by financial traders. Financial traders require a premium for holding LTTRs, which explains their systematic underpricing relative to forward markets. I show that inframarginal LTTR holders can lock in profits by taking forward market positions immediately after the auction, effectively engaging in arbitrage across LTTR and forward markets. Building on option pricing theory, I show that their aggregate forward market positions should be short in importing and long in exporting markets. Using 2018-2025 data, I find empirical support for this hypothesis in the German market. Arbitrage therefore attenuates but cannot eliminate the mispricing between transmission rights and forward markets. These results confirm that holders of transmission rights earn systematic rents. The corresponding losses are ultimately borne by consumers through higher grid fees, suggesting a transfer from consumers to transmission rights holders. These findings indicate that LTTRs deliver limited hedging value at substantial cost.
\par
This carries a lesson for transmission rights design. The fundamental problem with LTTRs is that TSOs are forced to sell insurance for which there is too little demand. Low demand for LTTRs as a hedging instrument is not only caused by the design shortcomings of LTTRs but also reflects that other products for hedging basis risk are traded on private markets. Any redesign of transmission rights should therefore rest on a thorough investigation of the incompleteness of these markets, since interventions are potentially costly. This also involves estimating the future demand for insurance against cross-border spread risk, given that both the build-out of physical interconnection capacity and the more efficient utilization of that capacity through flow-based market coupling can increase spot price convergence, reducing basis risk and thus the demand for insuring it through transmission rights or alternative products.

\clearpage

\subsection*{Data availability}

The data that support the findings of this study are available from the European Energy Exchange (EEX), but restrictions apply to the availability of these data, which were used under license for the current study and so are not publicly available. The code needed to reproduce the analysis is available under an open license at \url{https://github.com/cstiewe/lttr-arbitrage}.

\subsection*{Declaration of use of generative AI and AI-assisted technologies}
During the preparation of this work, I used Claude to assist with writing the code for the analysis. I take full responsibility for the content of the published article.

\subsection*{Acknowledgments}
I thank Lion Hirth, Jorge Sánchez Canales, Alice Lixuan Xu, Chiara Fusar Bassini, Vincent Thévenin, and participants at the Young Energy Economists and Engineers Seminar for valuable feedback and discussions, as well as EEX for providing access to the data used in this study. This work is supported by the German Federal Ministry of Research, Technology and Space (BMFTR) via the ARIADNE Project (FKZ 03SFK5K0-2). 

\clearpage
\appendix

\renewcommand{\thetable}{A\arabic{table}}
\renewcommand{\thefigure}{A\arabic{figure}}
\setcounter{table}{0}
\setcounter{figure}{0}

\section*{Appendix} \label{sec:appendix}

\begin{figure}[htbp]
  \centering
  \includegraphics[width=\linewidth]{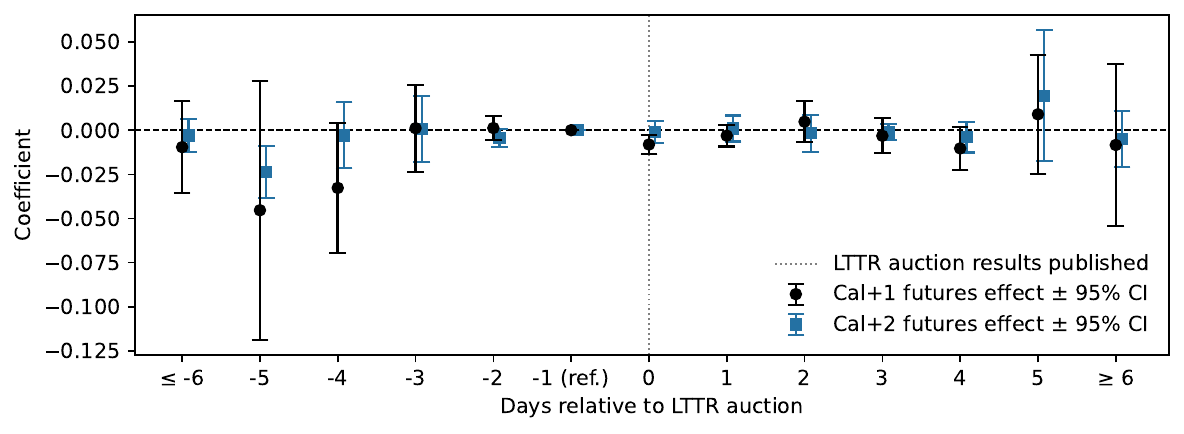}%
  \caption{Placebo test of the effect of LTTR auctions on second year-ahead German futures returns. This tests, for example, whether auctions for the 2025 LTTR affect not only Cal-2025 futures but also Cal-2026 futures traded on the same day.}
  \label{fig:event_study_nextcal}
\end{figure}

\begin{figure}[htbp]
  \centering
  \includegraphics[width=\linewidth]{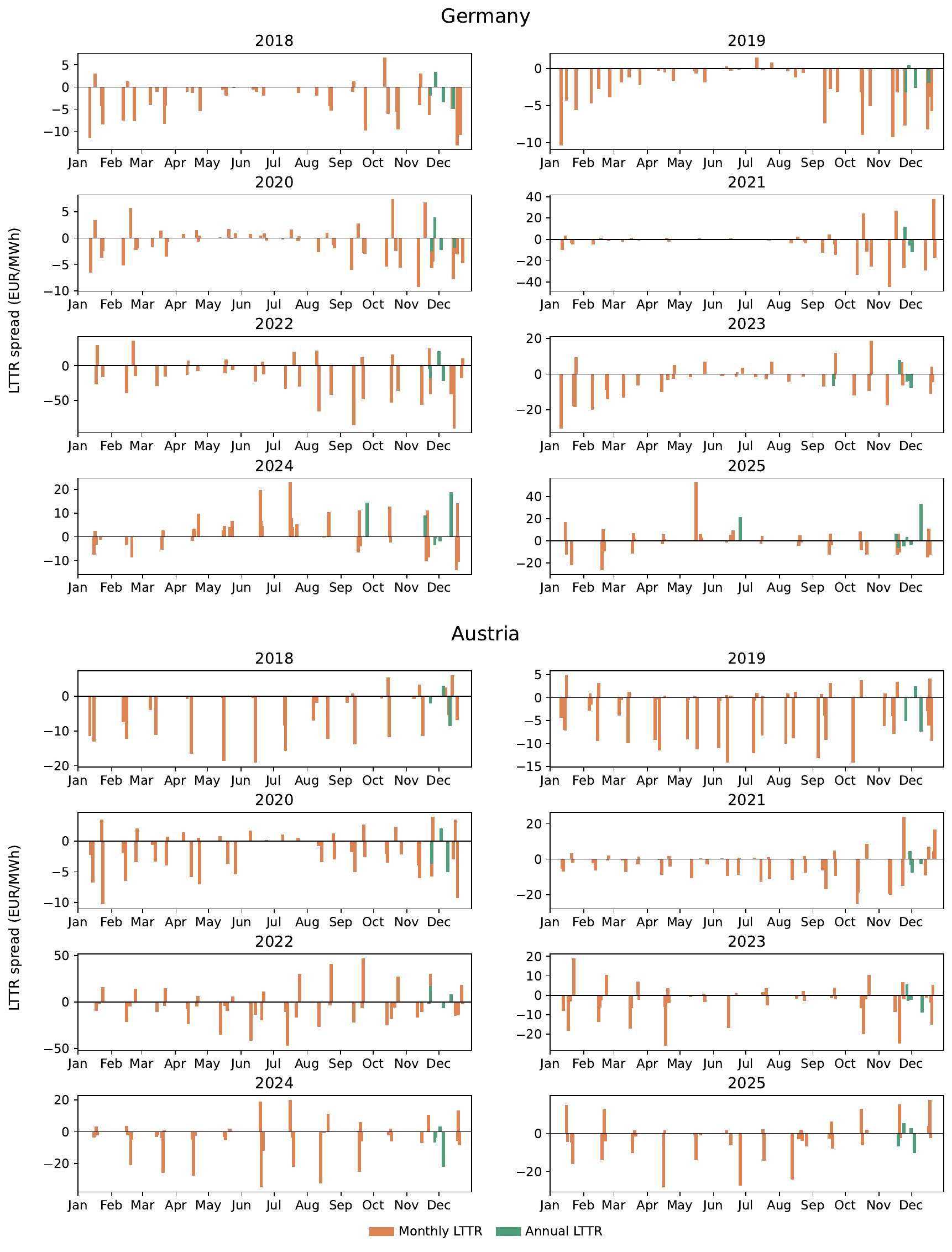}%
  \caption{Monthly and annual LTTR spreads, Euro/MWh. Bars indicate daily spreads in LTTR auction prices. Positive (negative) values imply that importing (exporting) LTTRs are valued higher.}
  \label{fig:lttr_spread}
\end{figure}

\clearpage
\bibliographystyle{apalike}  
\bibliography{xb_forward_markets}       

\end{document}